\documentstyle[12pt,epsf,colordvi]{article}                                    
\newcommand{\Figref}[1]{Figure~\ref{#1}}
\newcommand{\Cardiff}{{\em Department of Physics and Astronomy}
 \\{\em University of Wales College of Cardiff, Cardiff, U.K.}}
\newcommand{\Glasgow}{{\em Department of Physics and Astronomy}
 \\{\em University of Glasgow, Glasgow, U.K.}}
\newcommand{\MPQ}{{\em Max Planck Institute for Quantum Optics}
 \\{\em Garching, Germany}}

\begin{document}

\title{Results of the First Coincident Observations by Two\\
 Laser-Interferometric Gravitational Wave Detectors}

\author{D.~Nicholson, C.\,A.~Dickson, W.\,J.~Watkins, \\ 
B.\,F.~Schutz, J.~Shuttleworth, G.\,S.~Jones \\ \Cardiff \\ \\
D.\,I.~Robertson, N.\,L.~Mackenzie, K.\,A.~Strain, \\
B.\,J.~Meers, G.\,P.~Newton, H.~Ward, \\ 
C.\,A.~Cantley, N.\,A.~Robertson, J.~Hough \\ \Glasgow \\ \\ 
K.~Danzmann, T.\,M.~Niebauer, A.~R\"udiger, \\
R.~Schilling, L.~Schnupp, W.~Winkler \\ \MPQ}

\maketitle
\begin{abstract}

We report an upper bound on the strain amplitude of gravitational 
wave bursts in a waveband from around 800\,Hz to 1.25\,kHz. In an 
effective coincident observing period of 62 hours, the prototype 
laser   interferometric   gravitational   wave   detectors of the 
University of Glasgow and Max Planck Institute for Quantum Optics, 
have  set  a  limit of $4.9 \times 10^{-16}$, averaging over wave 
polarizations  and  incident directions. This is roughly a factor 
of  2  worse  than  the theoretical best limit that the detectors 
could  have  set, the excess being due to unmodelled non-Gaussian 
noise.  The experiment has demonstrated the viability of the kind 
of observations planned  for the large-scale interferometers that 
should be on-line in a few years time.

\end{abstract}

\newpage

\section{Introduction}
Gravitational   radiation   is   expected   from a wide range of 
astrophysical  sources  such  as stellar collapses,   mergers of 
neutron star and black hole binaries, pulsars, and from the very 
early  universe.  In  order  to  have  an  appreciable chance of 
detecting  this  radiation,  theoretical  calculations  indicate 
that  gravitational  wave detectors should attain  an  effective 
strain sensitivity  $h$   better  that  about $10^{-21}$  over a 
bandwidth from a few hundred Hz to about $1$\,kHz\cite{300yrs,texas}.   
It is  anticipated  that  this  target  will be reached  in  the 
next   few    years   by   large-scale   laser   interferometric 
detectors\cite{LIGO,VIRGO,GEO}.  A comprehensive   overview   of 
gravitational   wave   detection   can   be  found in two recent 
books\cite{blair,saulson}. 

At 15:02 (GMT) on 02 March 1989 two prototype gravitational wave 
detectors,  one  operated  by the University of Glasgow (UG) and 
the other by the  Max Planck Institute for Quantum Optics (MPQ), 
participated  in  a  joint  observing  run  over a period of 100 
hours.   The  motivations  for this run were twofold.  First, to 
demonstrate  the  practicality  of  making  long-term coincident 
observations  with  interferometers,  and second to provide real 
data  with all its inherent complexities for testing out a range  
of  signal  analysis programs.  This was the first time that two 
interferometers had been run in coincidence for such a length of 
time.  The noise performances  of  the  detectors during the run 
were  poorer  by  more  than  a  factor  of  ten  times what the 
prototypes could  achieve  today.   They might have been able to 
detect  a  nearby  (1\,kpc)  gravitational collapse event in our 
Galaxy,  the  probability of which in any 100-hour period may be 
between  $10^{-5}$ and $10^{-6}$.  We give brief descriptions of 
the detectors,  the  experiment, and  the results in this paper. 
Further details are given elsewhere\cite{robertson,dnprep}.

\section{The Detectors}
The  UG  detector  is  a  Fabry-Perot interferometer situated at 
latitude $55.86^\circ$~N, longitude $4.23^\circ$~W; its arms are 
orthogonal, at $193^\circ$ and $283^\circ$ clockwise from north. 
The  MPQ  detector  is  a  Michelson  delay-line  interferometer 
located  at   $48.24^\circ$~N,   $11.68^\circ$~E; its  arms  are 
rotated by angles of $31^\circ$ and $121^\circ$ from north.  The 
separation of the detectors on  a  chord  through  the  Earth is 
1370\,km,  corresponding to a wave travel time  (at the speed of 
light) of 4.6\,ms. A measure of the common  sensitivity  of  the  
detectors to impinging gravitational waves is provided by the 
overlap between the detectors antenna patterns. 
The  prototype  detectors  have  patterns  that  overlap by 89\% 
implying that their arms that are nearly aligned. As we shall see, 
this feature simplifies the coincidence analysis.  
 
\Figref{fig:noise}(a), displays the linear strain noise spectral 
density  of  the  UG  detector  at  a time when the detector was 
functioning  normally and close to optimally.  Above  about  1.5\,kHz,  
the  instrument  
operated near to  it's  shot-noise  limited sensitivity.  The 
excess  noise  below  this  frequency was  due mainly  to ground 
vibrations   and    mechanical   resonances.    The  $h$  strain 
sensitivity of  the   UG   detector,   in   a   frequency   band 
$0-1.25$\,kHz  which  we  motivate  later,   most  often  had  a 
measured value of around $2 \times 10^{-17}$. At times,  however, 
the  sensitivity  was observed to improve to a level $10^{-17}$,  
and conversely there were transitory spells when the sensitivity 
would  degrade by more than a factor of five times this value. 

The  linear  strain  noise spectral density curve for  the   MPQ  
detector,  again  during normal and near-optimal operation,    is    
shown    in 
\Figref{fig:noise}(b).    It  shows  features that  are  broadly  
similar  to  the   corresponding  curve  for  the  UG  detector.  
High-pass  and  low-pass  (anti-aliasing)  filters,  with corner 
frequencies of $320$\,Hz and $4$\,kHz respectively, were applied 
to the  data before  it was recorded. Their effects are illustrated 
by the faint line in the figure. Noise  peaks are  
evident at a frequency near $3.5$\,kHz. Their origin is known to  
lie  in  mechanical resonances  within the detector. The typical  
$0-1.25$\,kHz $h$ strain sensitivity of the MPQ detector  during 
the  experiment had a measured value around $3 \times 10^{-17}$. 
About   midway  through  the  experiment  this  sensitivity  was 
observed to degrade by more  than  a  factor  of two due to loss 
of alignment of the laser beams  in the vacuum tubes.   In common 
with  the  observed behaviour of the UG  sensitivity, there were  
also rare, transitory,  periods  when the sensitivity of the MPQ 
detector was  poorer by  more  than  a  factor of five times its 
typical measured value.   

\section{The Observing Run}
In  this  section we describe briefly the operational details of 
the   experiment.    Further  information can be found elsewhere
\cite{robertson,niebauer,dnprep}. 

The UG detector produced 60\,kB/s of data recorded onto 28 tapes  
by an  Exabyte tape drive. One third of this was the main output 
data stream, called  the  secondary error point signal. The rest 
was primarily housekeeping data. The whole set amounted to about 
20\,GB of data for the entire experiment. 

The  MPQ  detector  produced  about 44\,kB/s of data, which were 
recorded  onto  94  standard  6250~Bpi  9-track  tapes. The main 
interferometer signal was sampled at $10$\,kHz and compounded by 
several  housekeeping  streams. A total data set of about 15\,GB 
was amassed during the experiment.  

Interferometers are intrinsically more difficult to operate than 
bar detectors, since they involve a  number  of  active  control 
systems  that  must  constantly  be  monitored  and occasionally 
corrected.   Designs  for  the  large-scale interferometers will 
incorporate  many control features whose function is to minimize 
the need for  operators  to   intervene with the  running of the 
detectors.  The   present  prototypes  do  not incorporate  such 
features, since they  were  designed  as  scientific development 
test-beds rather than  as  observatories.  It was therefore very 
encouraging that the  prototypes   performed   so   well:   they   
acquired   data  simultaneously during  88\% of  the experiment, 
and operated close to their  optimum  sensitivity simultaneously  
for  62\% of the experiment\cite{dnprep}. 

\section{The Analysis Method}
Several  of  the  general  issues  that  arise  in  the analysis 
of  gravitational  wave  data  have  been  addressed  in  recent 
monographs\cite{gwda,bfsblair}.   A   limited  analysis  of  the 
UG data has been performed\cite{robertson}, and the MPQ data set 
has been searched for a pulsar signal\cite{niebauer}.  Our focus 
here is  on  short-time-scale  bursts  that could be produced by 
supernova explosions.  Subsequent  publications will analyse the 
same data to place upper  limits  on  the gravitational wave flux 
of periodic waves  impinging from  a  small solid angle on the 
sky\cite{gsjthesis}, and of a stochastic background\cite{kacthesis}. 

Our analysis method was basically  to  compute cross-correlation 
integrals for each individual data stream in turn, with a filter 
weighted by the inverse  of  the  power  spectral density of the 
noise  for  that  data  stream.  This is the classical method of 
matched-filtering\cite{wainstein}.   The  inverse noise spectral density 
weighting in the filter supresses  contributions to the integral 
from   frequency  bands  where  the data is very  noisy.  It was  
important,    of course,    to   remove   the anti-aliasing  and 
high-pass  filters  from  the  MPQ  data  to  ensure  that  this 
weighting would be done properly. 

The Fourier transform of the filter we employed is  given simply 
by   $\Pi(f/2500\;\rm\,Hz)$   where   $\Pi(x)$  is  the ``boxcar
function'' \cite{bracewell},  of  unit value  for $|x|<1/2$  and
zero otherwise.  The filter's output is thus the inverse Fourier 
transform  of  the  noise-weighted  data   stream   using   only 
frequencies below the cut-off of 1.25\,kHz.  It  is  evident from 
\Figref{fig:noise} that the  detectors are  most  sensitive in a 
frequency  range  from  about  $800$\,Hz  to  $3$\,kHz, and this 
overlaps with only $1/5$ of the bandwidth of our  burst  filter.  
However,  the performance of resonant bar detectors is generally
quoted in terms of their sensitivity  to flat spectrum bursts in 
the  approximate  frequency  range $0-1.5$\,kHz, even though bars
are only sensitive in a  much  narrower  band\cite{barref}.  Our 
burst filter was therefore chosen  for reasons  of compatibility 
with the model  adopted by the resonant bar community,  in order  
to   facilitate comparisons  between respective sets of results. 
Some  numerical   simulations   have   also   reinforced     the  
expectation   that   supernova  bursts  will be broadband in the 
kilohertz region\cite{Finn}.  

\section{The Analysis Software}

The  100-hour  experiment  was undertaken with a view to gaining 
practical  experience  under  realistic  conditions and offering
guidance for the development of larger detectors. This motivated
our  decision to  analyze the  data with  computer programs that 
could  serve as prototypes of programs that will have to process 
the  data  of  the  large  interferometers  in  real  time. This 
software, designed by the Cardiff group, is described  in detail 
in a Ph.D. thesis\cite{watkins}. 

\subsection{Identifying coincidences}
The programs produced lists of ``events''.   An event is defined 
as a set of  contiguous  data  samples  where  the output of the 
filter exceeds a predetermined threshold.  The threshold was set 
low  enough  to  generate  a reasonable number of events for the 
statistics  of  the  coincidence search. Several diagnostics for 
each event were recorded onto Exabyte tapes, including the time, 
duration,   and  maximum  amplitude  of  the event,  and numbers 
characterizing the detector, such as  the broad-band sensitivity 
to  kilohertz  bursts  in short time slices of filter output and 
various  housekeeping  indicators  (microphone  and  seismometer 
signals etc.) at the time of the event.

A coincident event is defined to be a pair  of  events  from the 
two data streams that occur within  the  acceptance window given 
by the gravitational wave travel time between the detectors. The 
list of coincidences  inevitably  contains  some with very large 
amplitude,  and  one  has  to  assess their significance.  It is 
important  to  formulate {\it a priori} criteria  for  accepting 
events, without reference to the specific properties of the data 
set.  Our analysis proceeded through 3 levels of vetoes.

At  the  first level, we looked at housekeeping data and applied 
vetoes if the detectors were not performing correctly,  or if an 
environmental  disturbance  had occurred which may have affected 
the detector.  The vetoes  were  partially,  but  not  entirely, 
successful  in  eliminating  periods  when  the  detectors  were 
performing below par.  There  were  unvetoed  times  during  the 
experiment  when  the kilohertz   burst   sensitivities  of  the 
detectors were 5 times larger than their typical measured values.

The second level  accepted  only data  from  periods  where  the 
sensitivity of the two detectors was  nearly  optimum.  At other 
times,  the information  from  the  experiment  was less useful. 
For around 20\% of the time that the UG detector took data, it's 
kilohertz burst sensitivity exceeded  $4 \times 10^{-17}$. 
Similary, for 20\% of it's observing time the MPQ detector had 
a kilohertz burst sensitivity in excess of $5 \times 10^{-17}$. 
A  decision was 
made to reject coincidences if the events  from  the  individual 
streams belonged to times when the  filtered  strain noise  fell 
into the respective 20\% tails. This  reduced  the  effective 
coincidence observing period of the detectors to around 62 hours. 
  
At  the  third  level, events that survived the first two vetoes 
were  tested  against  a  simple  model:  if  they  were genuine 
gravitational waves, they ought to have shown the same intrinsic 
strain  amplitude  in  both detectors, apart from the effects of 
detector  noise. This is by virtue of the  near-alignment of the
interferometers arms which was noted above.  We  calculated  the 
probability  that  the two observed strain amplitudes could have 
been produced  by  an   unknown  signal  plus  independent noise, 
given  the noise level in each  detector  at  the  time  of  the 
coincidence, and based on a  model of independent Gaussian noise 
in the two detectors. If this probability was  less  than  0.1\%,  
we  rejected  the  coincidence.  The  reason  for rejecting such 
events  is our  {\em  assumption} that  any  real  gravitational  
wave   events of  this  strength  will  be  very rare, and it is 
therefore unreasonable to conclude that a coincidence  is caused 
by a gravitational wave if to do so requires  us to assume that, 
in addition, the detectors were behaving in a  very  unusual way 
as well when that rare event occurred. We call this the $h$-veto.

Coincidences that  survive the three levels of rejection set the 
upper limit on the sensitivity of our experiment.

\section{Results}
We chose a threshold of $4\sigma$ for generating events from the 
filters.   The  top  panel  of  \Figref{fig:results}  shows  the 
signal-to noise ratio (SNR) distribution  of  coincident  events 
that  pass  the  housekeeping  vetoes  during  good data periods 
(levels 1 and 2).  A  calculation  based  on  the  empirical SNR 
distributions  of  {\sl all}  the events  in  each detector that 
survive  these  vetoes,  yields  a  probability  for  the  least 
likely coincident event in this diagram of 0.51, so none of  our 
events are statistically significant outliers. The bottom  panel 
of the same figure shows events that pass the $h$-veto,  so that 
they could have been generated by a  gravitational  wave  with a 
reasonable amount of noise on top.   The axes here are in strain 
rather than SNR.  The inferred intrinsic amplitude of coincident 
bursts is the average value of the strains on the two axes.   We 
conclude that our limit on  kilohertz  bursts  of  gravitational 
radiation is $2.2 \times 10^{-16}$.  This  is  the  limit if the 
waves arrive from the zenith with the optimum  polarization. The 
corresponding limit on bursts  of  any  wave polarization coming 
from  any  direction  on the sky is a factor of $\sqrt{5}$ worse, 
or $4.9\times 10^{-16}$.

\section{Discussion}
Our  limits  are  the  first obtained over a broad gravitational 
wave bandwidth.  The false-alarm  threshold  for  a single alarm 
during the effective coincidence observing  period,  taking into 
account  the   light  travel-time  between  the  detectors,  and 
assuming a  background  of  independent  Gaussian  noise  in the 
detectors, is $4.5\sigma$.  Given  the  typical  kilohertz burst 
sensitivity of the detectors,  we  estimate that our upper bound 
on  $h$  is  only  about  a  factor of  roughly 2 worse than the 
theoretical best  limit that these detectors could have set. The 
most recent result  that  has  been  published in any detail for 
coincident experiments  between  resonant bars, has set an upper 
limit  on   the  amplitude  of  kilohertz  bursts  of   $1\times 
10^{-17}$\cite{barref}. This  is  for a narrow  waveband however 
(and  it  is  not stated whether  this  limit  incorporates  the  
sky-averaging  factor).   Interferometer  prototypes  have  been 
markedly improved since 1989, and would probably come very close 
to this limit if a similar experiment to the 100 hour  run  were 
performed today. However, the real  value  of  our  results is a 
test  of   interferometric   observing.  Our  results  are  very 
encouraging for large-scale interferometers, since they indicate  
that  attention  to  detector  control  and  non-Gaussian  noise 
could raise the sensitivity and duty cycle of working  detectors 
very close to their optimum performance.

\newpage

\begin{figure}\caption{
Linear noise spectral density curves for the (a) University of 
Glasgow  (UG) and  (b)  Max  Planck  Institute  for  Quantum Optics (MPQ)  
prototype    gravitational    wave   detectors  under   normal 
and   near-optimal operation during the  100-hour  experiment.  
The faint line in  curve   (b)   illustrates  the   effect  of 
anti-aliasing and  high-pass   filters  applied  to data  from 
the Max Planck detector, which were removed for the data analysis.
\label{fig:noise}
}\end{figure}

\begin{figure}\caption{
A signal-to-noise ratio scatter plot (a) of coincident events passing the 
first two levels of analysis and a strain amplitude scatter plot (b)
for events that also pass the $h$-veto. See text for details. 
\label{fig:results}
}\end{figure}

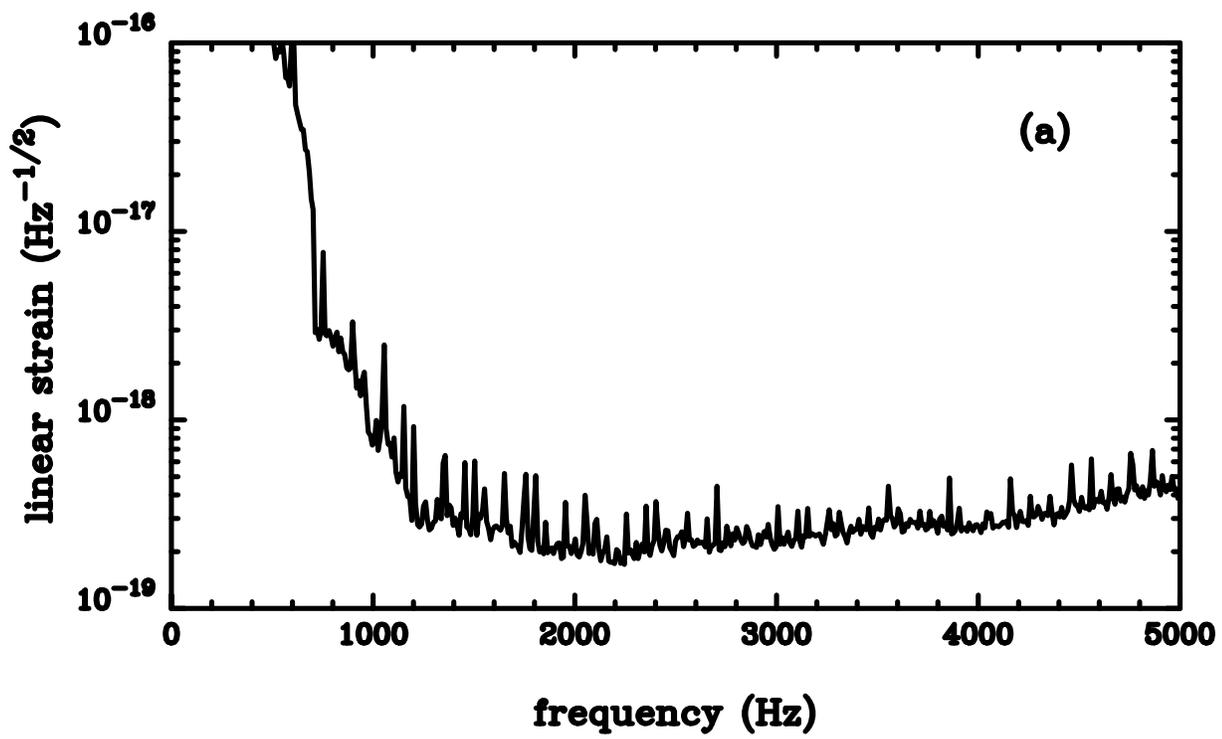
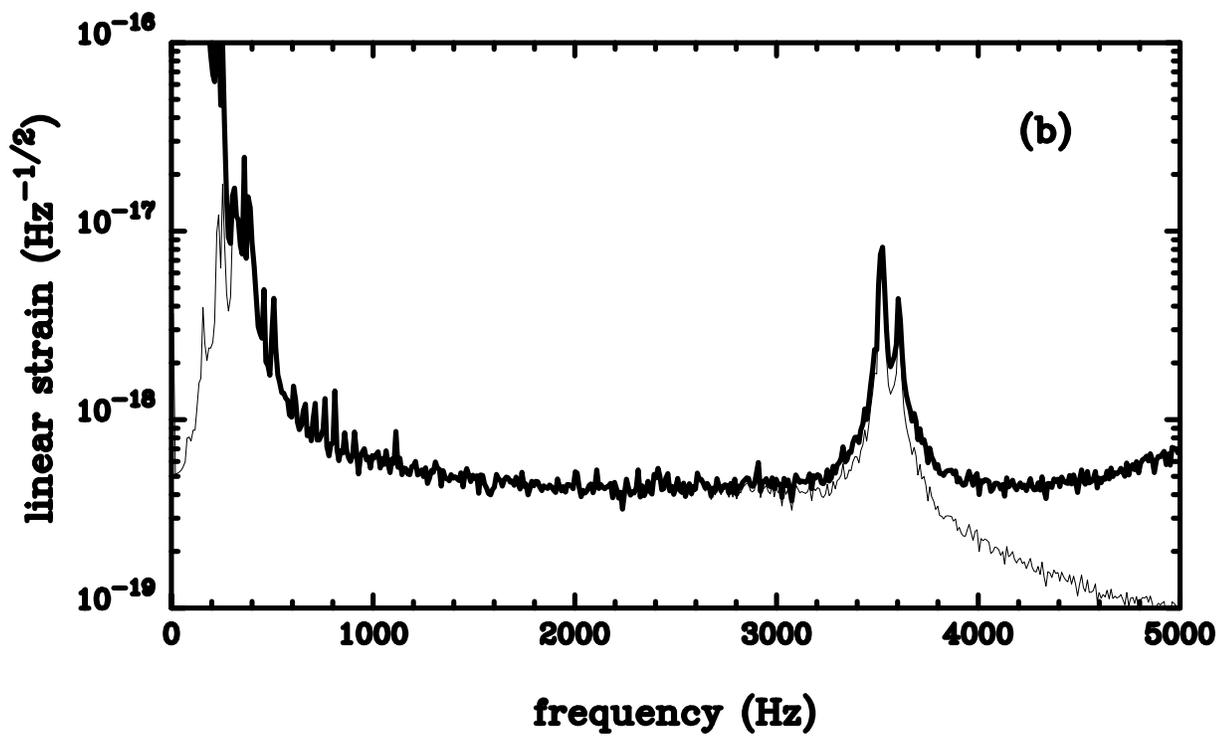
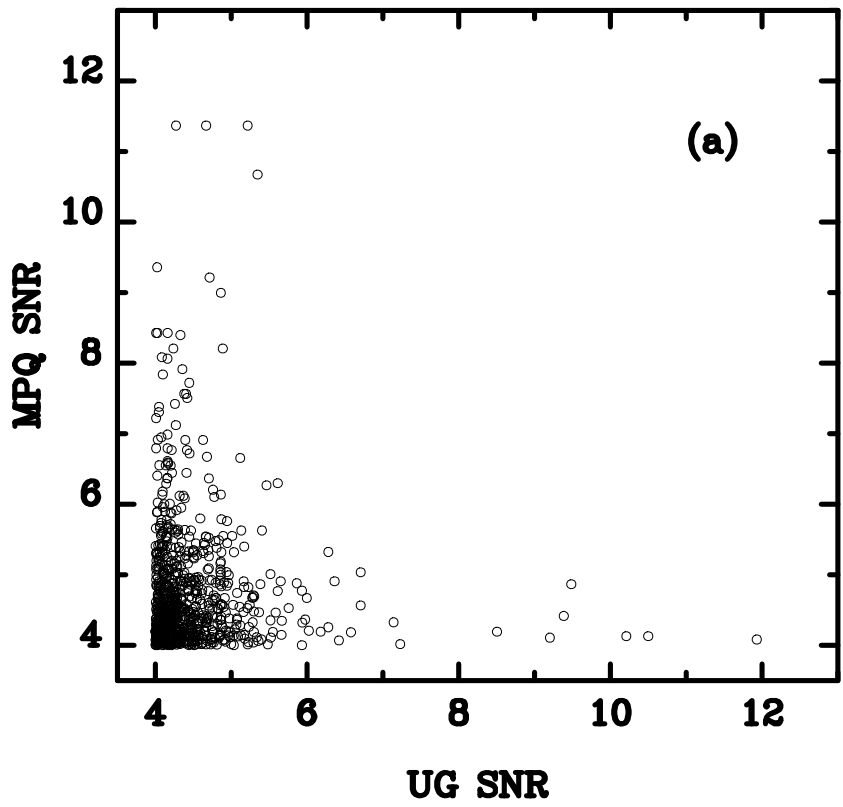
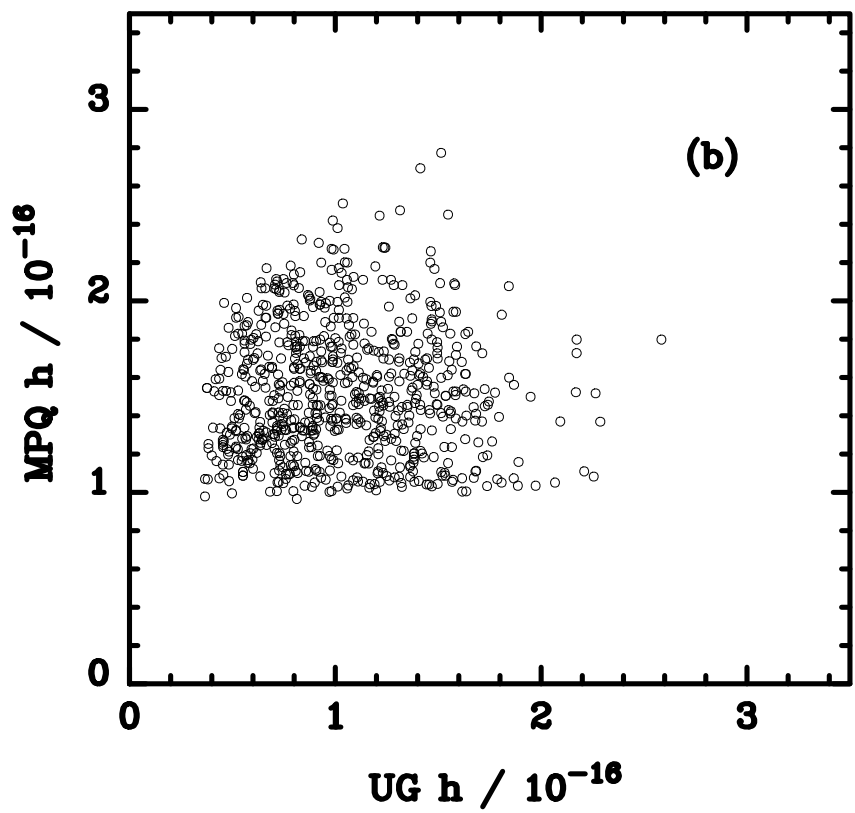
\end{document}